# Spatial evolution of Hindmarsh–Rose neural network with time delays

Michał Łepek*, Piotr Fronczak

Warsaw University of Technology, The Faculty of Physics, Koszykowa 75, 00-662, Warsaw, Poland

*lepek@if.pw.edu.pl

***Abstract.*** Spatial relations between neurons in the network with time delays play a crucial role in determining dynamics of the system. During the development of the nervous system different types of neurons group together to enable specific functions of the network. Right spatial distances, thus right time delays between cells are crucial for an appropriate functioning of the system. To model the process of neural migration we proposed simple but effective model of network spatial evolution based on Hindmarsh–Rose neurons and Metropolis–Hastings Monte Carlo algorithm. Under the specific assumptions and using appropriate parameters of the neural evolution the network can converge to the desirable state giving the opportunity of achieving large variety of spectra. We show that there is a specific range of network size in space which allows it to generate assumed output. A network or generally speaking a system with time delays (corresponding to the arrangement in the physical space) of specific output properties has a specific spatial dimension that allows it to function properly.

## 1. Introduction

Complex nervous systems, such as human brain consisting of about 10 billion neurons [1], are the systems that most of important questions related to, remain unanswered. Neurons integrate signals and encode information and although activity of a single nerve cell can be well–modeled today, the phenomena occurring in complex biological neural networks still need to be studied to understand the mechanisms behind various brain functions.

In recent decades, many works focused on the synchronization between neurons [1–5] as this mechanism was believed to be vital for several cognitive functions. The concept of small–world networks by Watts and Strogatz was also included [1, 6]. The recent innovation is the introduction of time delays to network models [7–23] which allows to study the influence of the limited speed of information processing on the network dynamics. This speed limitation is indeed present in neuronal communication as the action potential propagates with the speed of tens of meters per second which is a significant aspect if the physical size of nerve tissue taken into consideration. There are various sources of information time delay between neurons, e.g. limited speed of transmitting action potential through the axon, different types of synapses (chemical, electrical), the release of neurotransmitter, the condition of myelin sheath. Time delays are the important parameters influencing the formation and transitions between neuronal firing patterns [12–14]. It has been shown that time delay drives the collective behaviors of neurons in the network [15]. Applying time delay to the network can enhance its synchrony or stop it from synchronizing [12–14, 18–20]. It also changes the bifurcation image of neural activity. Appropriate choice of time delays and coupling strengths allows to observe intermittent behavior of neurons [19] or spiral waves propagating in the network [22].

Recent papers on modeling dynamical systems including experimental reference to biologic neural networks also deserve interest. In [24] authors compare their numerical simulations to recordings of hippocampal place cells of rats. Simultaneously, using artificial neural networks for information processing it is possible to predict behavior of dynamic biologic system straightforward, not through modeling the system itself [25]. On the other hand, it is known nowadays that during early stages of development of the nervous system the process of neuronal migration occurs. As the positioning of nerve cells constrains local neuronal signal, the migration leads to grouping of different classes of neurons together providing them appropriate spatial relationships and thus ability of the appropriate interaction [26]. Recently, it has been shown that mechanism not covered in the field before are also important factors when neuronal behavior considered, e.g. autapse connections and even physical effects as electromagnetic induction and radiation [17].

What we propose in our paper is to combine two main concepts: time delays corresponding to positioning neurons in the physical space and their space–evolution in time based on the Metropolis–Hastings Monte Carlo algorithm [27]. In opposition to our predecessors we do not set constant distribution of time delays but allow them to vary as the nodes move on the surface due to the widely known and simple Markov chain Monte Carlo rules. We use Hindmarsh–Rose neuron model with connection topology of a regular ring lattice which is in general independent from the arrangement of neurons in the physical space. We show that taking a specific target function and appropriate simulation parameters it is possible for the system to evolve from one specified state to another. As a specified state we consider the power spectrum of the output from the network. The networks evolving in our study generate power spectra consisting of the chaotic and periodic parts. In particular, the network can evolve from the state of entirely chaotic output signal to the state of entirely periodic signal, generating a large variety of spectra. We also study physical extent of evolving network to answer the question whether there is a specific physical size that allows it to function in the pre–assumed way.

The paper is organized as follows. Neural network model with time delay coupling is introduced in Sec. 2. In Sec. 3 we describe the model of the spatial evolution and the target function. Details of numerical studies and results are given in Sec. 4. In the last Sec. 5 we briefly present conclusions and possible extensions to this study.

## 2. Neural network model

A single Hindmarsh–Rose neuron model can be represented with three differential equations, which are as following [28]:

$$\begin{cases} \dot{u} = v - au^3 + bu^2 - w + I & \text{(1a)} \\ \dot{v} = c - du^2 - v & \text{(1b)} \\ \dot{w} = r[\beta(u + \chi) - w] & \text{(1c)} \end{cases}$$

where $u$ is the membrane potential of the neuron and $v$, recovery variable, is related to the fast ion current (as Na⁺ or K⁺). Adaptation current is represented by $w$ (e.g. Ca⁺). The equations for $u$ and $v$ control the fast dynamics and the equation for $w$ controls the slow dynamics of the model. $I$ is an external stimulus which drives the activity of the single neuron. The model is able to produce a large variety of signals observed in biological nerve cells, as chaotic spiking, periodic spiking and bursting discharges. As many other authors [2, 3] for the numerical simulations we use: $a = 1.0$, $b = 3.0$,

$c = 1.0$, $d = 5.0$, $r = 0.006$, $\beta = 4.0$, $\chi = -1.56$. Time series of variable $u$ for different driving current regimes and the bifurcation diagram of a single neuron are presented in Fig. 1.

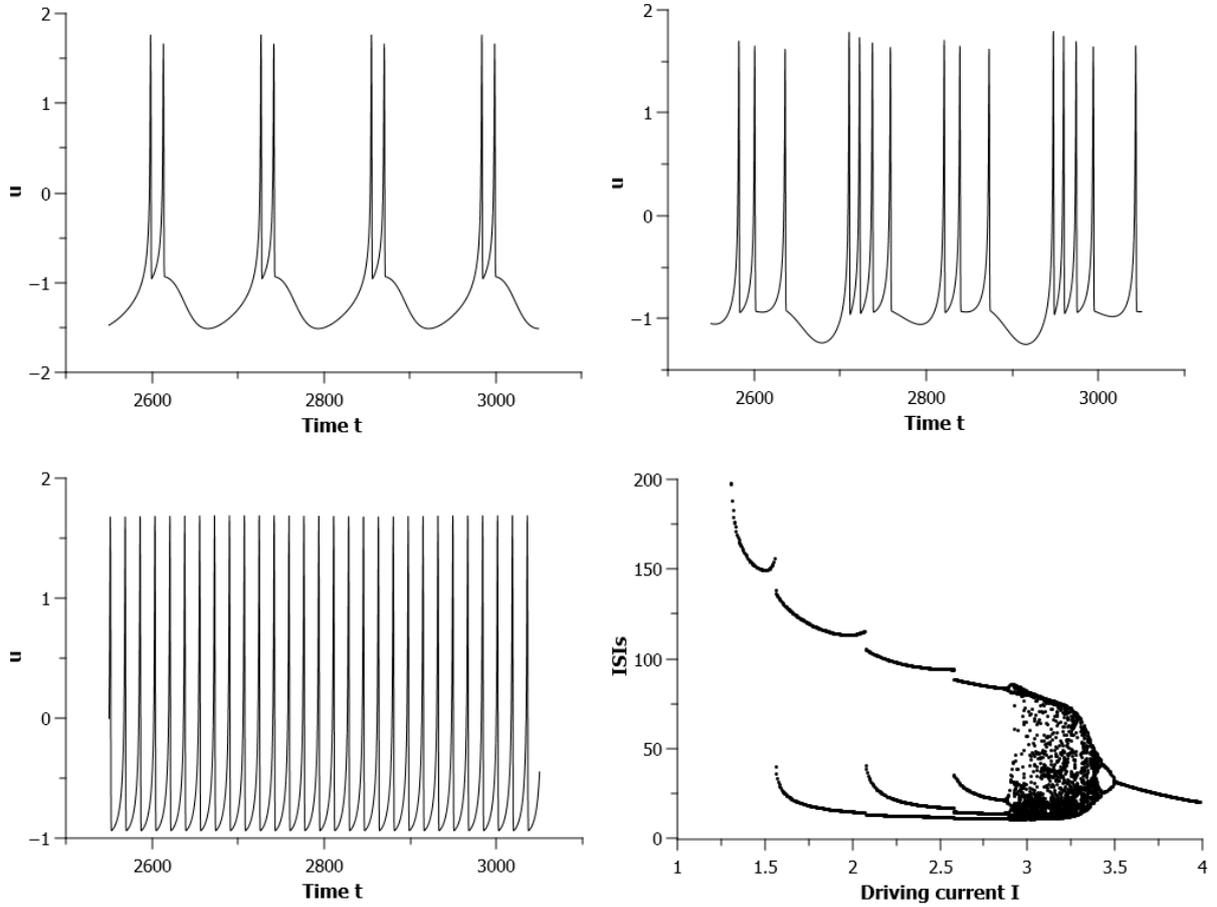

**Fig. 1.** Time series of variables u of a single neuron for different regimes. Upper–left: periodic regime, $I = 2$; upper–right: periodic regime, $I = 3.2$; lower–left: periodic regime, $I = 4.2$; lower–right: bifurcation diagram of inter–spike intervals (ISIs) versus driving current $I$.

Neural network can be represented as a set of $N$ connected nodes, each of them containing the model (1). The coupling between the nodes is realized by the differential components with time delay present that corresponds to the arrangement of the network in the physical space, e.g. on the plane surface. The neural network model is then described by the following set of differential equations [10, 28]:

$$\begin{cases} \dot{u}_i = v_i - au_i^3 + bu_i^2 - w_i + I_i + F_i(t) & \text{(2a)} \\ \dot{v}_i = c - du_i^2 - v_i & \text{(2b)} \\ \dot{w}_i = r[\beta(u_i + \chi) - w_i] & \text{(2c)} \end{cases}$$

where the effect on $i$ of coupling from $j$ is given by:

$$F_i(t) = k \sum_{j}^{N} g_{ij}[u_j(t - \tau_{ij}) - u_i(t)] \quad (3)$$

where $i, j = 1, 2, \ldots, N$ and $k$ is the coupling strength; $(g_{ij})_{N \times N}$ is the symmetrical ($g_{ij} = g_{ji}$) connectivity matrix, so if a link between $i$ and $j$ exists then $g_{ij} = 1$ and if not then $g_{ij} = 0$. The neuron $i$ receives the signal from the neuron $j$ after the time of $\tau_{ij}$, i.e. $\tau_{ij}$ is the time delay that only depends on the spatial distance between the nodes. The time delay length is given by:

$$\tau_{ij} = \tau_{ji} = \text{int}[pd_{ij}] \Delta t = \text{int}\left[p\sqrt{(i_x - j_x)^2 + (i_y - j_y)^2}\right] \Delta t \qquad (4)$$

where int[·] stands for integer part, $\Delta t$ for time step, $d_{ij}$ is the distance between neurons $i$ and $j$, $p$ is a scaling factor (in this work we use $p = 13$) and $(i_x, i_y)$, $(j_x, j_y)$ are the coordinates of neurons. Time delays in the network we consider are selected in a way that corresponds to the arrangement of the network on the plane surface, so $d_{ij}$ denotes the Euclidean distance.

## 3. Evolution model

To study the spatial evolution of the network we worked out the following concept. We choose few of the neurons from the network to be output neurons and collect the signals they produce. The reason behind selecting few neurons instead of all of them is the observation that in the vast majority of complex neural networks (both applied information processing and biological models) not all of the nodes contact with external environment [29, 30]. Then, using Fast Fourier Transform the power spectrum density of the sum of the signals is computed. The power spectrum density function was chosen as it can be understood as a single output from the group of nerve cells of specified type and, simultaneously, it contains the information on different signals present in the output of this group. Due to the network arrangement on the surface, thus due to the various sets of time delays, the neurons can produce different signal patterns, therefore different power spectra can be observed. The examples of these are presented in Fig. 2. Generally speaking, in the model described above we observed that a neuron tends to produce a chaotic signal if most of its connections are short–distance (distance of ~1 unit in our model) or very–long–distance (roughly, distance of two orders of magnitude longer, ~100 units). A periodic signal can be observed if most of the connections are of the length between the two states mentioned above (long–distance, an order of magnitude longer than 1 unit). The relation between the distance from other neurons and the signal produced is shown in Fig. 3. It is worth to be emphasized that neurons that are not chosen as output neurons also have important impact on the output spectrum. Moreover, notice that the neurons in the network we study are in desynchronized state.

It is now possible to arrange a network where output neurons produce signals from different regimes (chaotic, periodic), so as the output of the network we have a specified (target) power spectrum. Then, we start with a new network, where all neurons are located in one, small–spanned group on the surface. To change the locations of the neurons and therefore time delays of their connections we use the Metropolis algorithm. During each iteration of the spatial evolution we randomly select a node and move it slightly on the surface. If the resulting position causes that current power spectrum fits the target spectrum better, we accept the new position; if not, we accept it with probability $P = \exp(-dE/T)$, where $dE$ is the change of the target function value and $T$ is the temperature. To measure the similarity of power spectra (target function) we use Pearson correlation coefficient. As the correlation coefficient takes values from the range of $[-1, +1]$, we put the target function as $E(t) = 1 - corr(S_{current}(t), S_{target})$, where $S_{current}(t)$ and $S_{target}$ stand for current power spectrum and target power spectrum, respectively. The evolution is stopped when stopping criteria are satisfied (e.g. low value of target function $E$).

In our simulations the search in physical space is enriched by varying the driving current $I_i$ of the neuron $i$ that is currently moving. Thus during each evolution iteration we randomly choose a spherical vector from the space $(x, y, I)$ of two physical dimensions and one current dimension. This additional feature is introduced to obtain a larger variety of possible output states of the system.

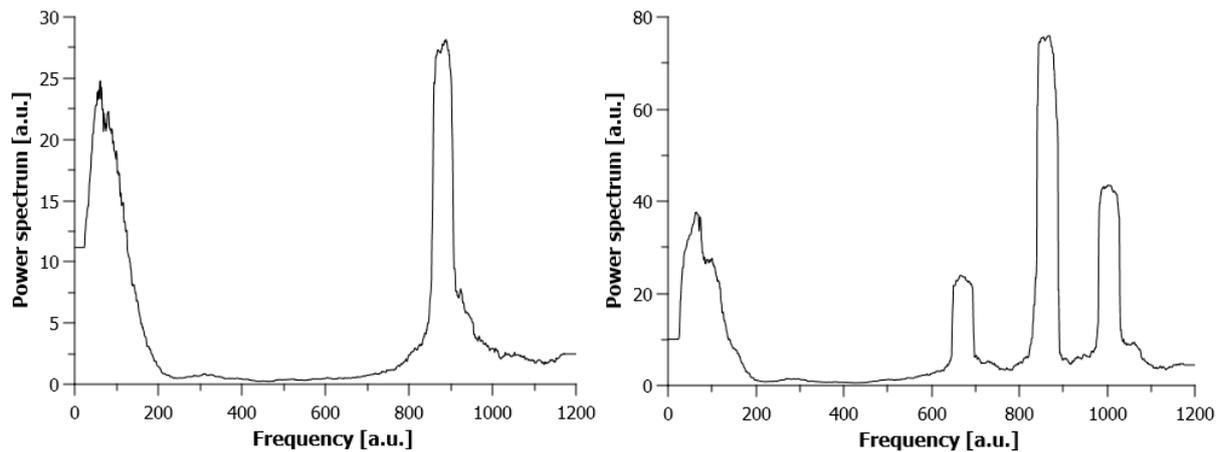

**Fig. 2.** Examples of different power spectra. The wide peak on the left of each diagram (0 ÷ 200 a.u.) is a result of the chaotic output neurons and peaks on the right–hand (600 ÷ 1100 a.u.) are result of the periodic output neurons.

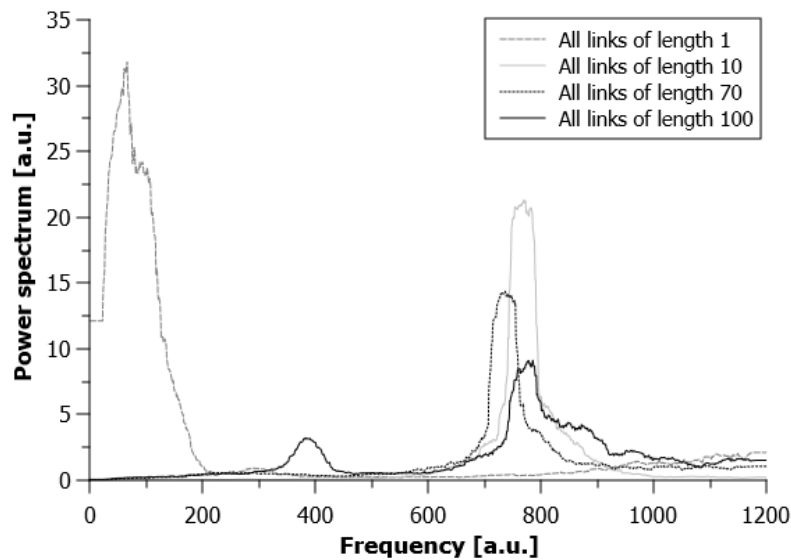

**Fig. 3.** Relation between the distance from other neurons and signal produced. When the output neuron is located close to the other neurons (all links of the length of 1) it produces a chaotic signal. When proper distance applied (all links of the length of e.g. 10), then signal turns to periodic one defined by driving current $I_i$. If further increase applied to the distance the signal starts to degenerate.

## 4. Numerical studies

One of the main obstacles to study the system is computational load which increases with the size of the network as the coupling $F_i(t)$ based on time delays is different for all of the neurons. Although we use highly optimized code, due to this issue we set a network of 10 neurons; as we will see later this number is enough for observing interesting phenomena. To check the robustness to the network size we also performed the simulation of network of 20 neurons.

The connectivity matrix so the connection topology is a regular ring lattice. Notice, that we do not specify time delays of the connections as they will be clearly defined by the initial positions of nodes, randomly selected on the planar surface later. The next vital preliminary question is the size of node neighborhood in the network. Fig. 4 shows target spectra for different values of neighborhood $s$. For small values of $s$ (e.g. $s = 2$) the network is very sparse, thus, for node $i$ the effect $F_i(t)$ from other neurons is very weak and output neurons are driven mostly by their driving currents $I_i$. For high values of $s$ (e.g. $s = 10$) the network is dense, so the strong effect $F_i(t)$ causes that dynamics of all the output neurons switches to chaotic regime. For further studies we apply the moderate value of $s = 4$ as it allows both chaotic and periodic output signals to co–exist in one network.

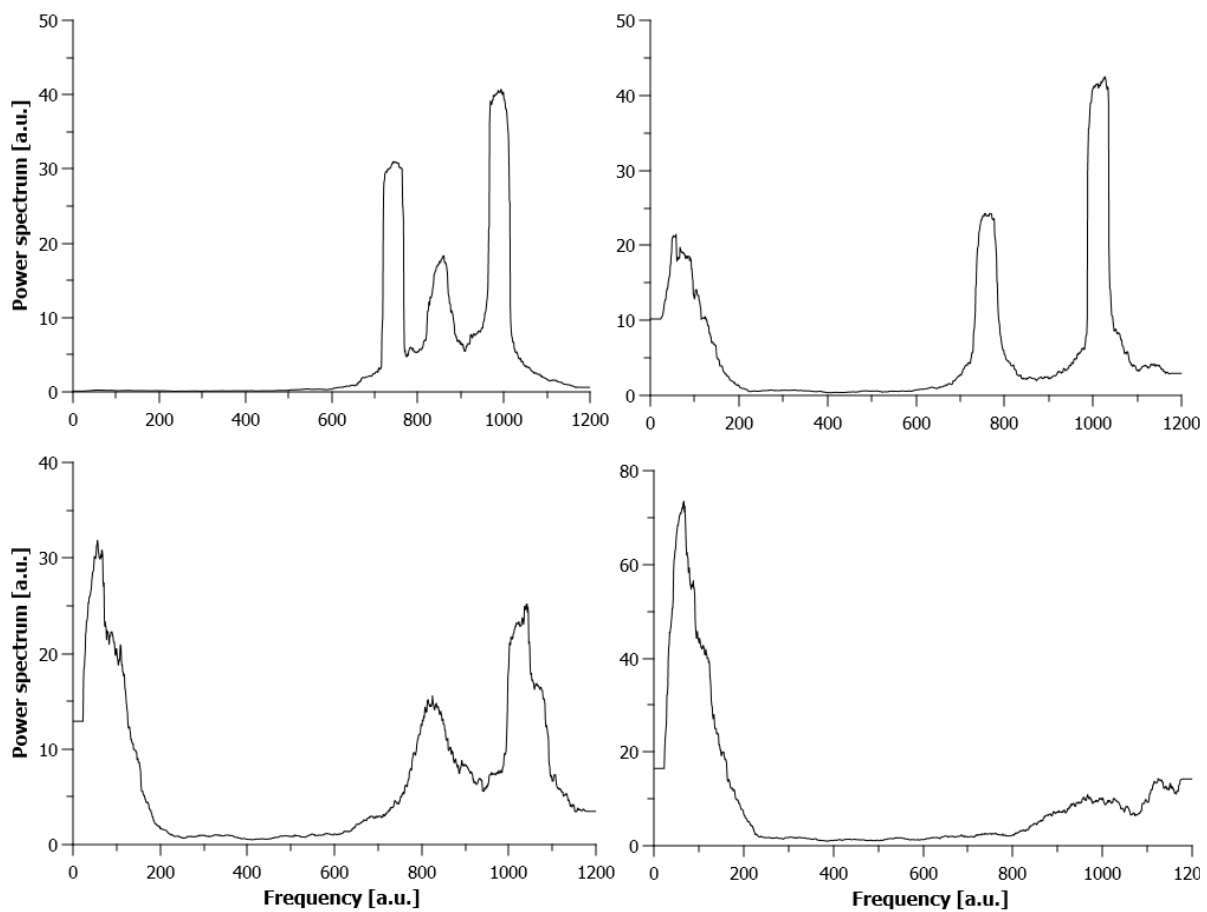

**Fig. 4.** The spectra for different values of neighborhood $s$. Upper–left: $s = 2$, upper–right: $s = 4$, lower–left: $s = 6$, lower–right: $s = 10$. For further studies we apply the moderate value of $s = 4$ as it allows both chaotic and periodic output signals to co–exist in one network.

As the output nodes we choose 3 of 10 neurons, specifically 1–st, 4–th and 7–th neuron. As $s = 4$, each $i$–th node is connected to the nodes $(i-2)$–th, $(i-1)$–th, $(i+1)$–th and $(i+2)$–th, so the selection is to prevent output neurons to be linked directly what could disturb network output adaptability.

In following numerical studies we use the fourth–order Runge–Kutta algorithm with the time step of $\Delta t = 0.01$. This time step is short enough to preserve the neuron signal but also long enough to shorten the computational time of the spatial evolution to acceptable level. Variables $u_i$, $v_i$, $w_i$ are initially set to random values. The total integration time of each simulation run is 20000. The signals from the last 13000 time units are used to produce the output spectrum of the network (as described in Sec. 3.). Then first 1200 samples of spectrum (as it fully covers the region on interest) are smoothed with the moving mean with window of 48 and used to calculate the correlation coefficient.

The other parameters we use in the studies are as follows. The general coupling strength is $k = 0.044$. The spatial steps of the $(x, y, I)$–space evolution are depended on $E$ and $\Delta x, \Delta y \in [-4E, +4E]$, $\Delta I \in [-0.02, +0.02]$. The real step in each iteration of the evolution depends on the random spherical vector as mentioned in Sec. 3. The temperature during the evolution is $T = 0.02$ and after achieving the criterion $E < 0.04$ it becomes $T = 0.005$ to stabilize the system at the final state. Now, having set all parameters for Metropolis time evolution, we put the target spectrum of output neurons as shown in Fig. 4 (chaotic and two periodic peaks).

At the beginning of the simulation the nodes of the network are grouped on the plane surface with random locations (from the range of $x_{start}, y_{start} \in [0,1]$) and random driving currents $I_i \in [3.8, 4.6]$ (corresponding to periodic regime for a single neuron). For the beginning state of that kind only the chaotic peak is present in the spectrum, so the value of $E$ is relatively high. Note that the connection topology (ring lattice) is constant during the evolution and the only parameters that evolve are space positions and driving currents of neurons. The exemplary initial state of the network to be evolved is presented in Fig. 5. During the evolution the network spreads on the surface allowing some nodes to move away from the others and therefore changing their time delays. As Metropolis algorithm mainly selects better states, $E$ decreases in time (as shown in Fig. 6). The process often finishes after a finite time with the value of $E$ relatively close to 0, which corresponds to convergence of the network spectrum and the target spectrum (as shown at Fig. 7). The process of evolution is presented in Fig. 8. The spatiotemporal patterns of neuron signals at various stages of evolution are shown in Fig. 9. Action potentials of one of the output neurons changing during the evolution are presented in Fig. 10.

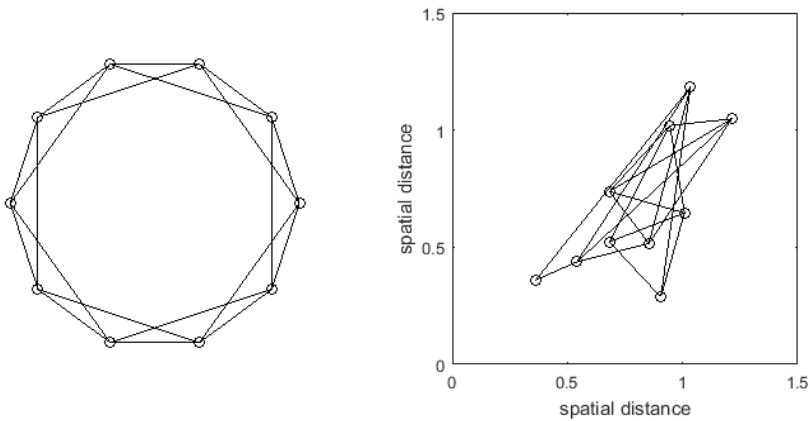

**Fig. 5.** Ring–lattice network connections with neighborhood $s = 4$ (left; the ring is only for better visualization of links) and the exemplary initial spatial state of neurons on the planar surface (right; the real spatial "look" of the network). The arrangement on the surface clearly defines the time delays of all of the links.

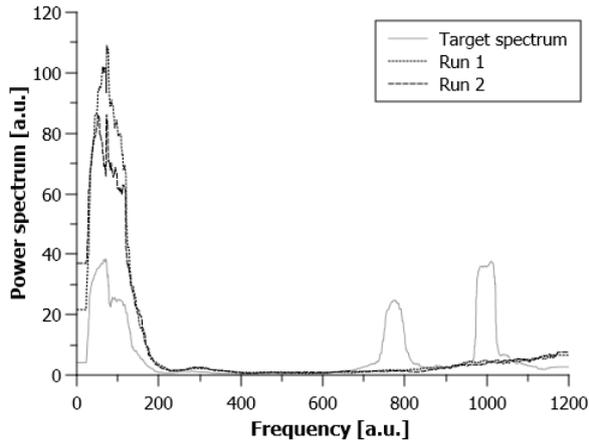 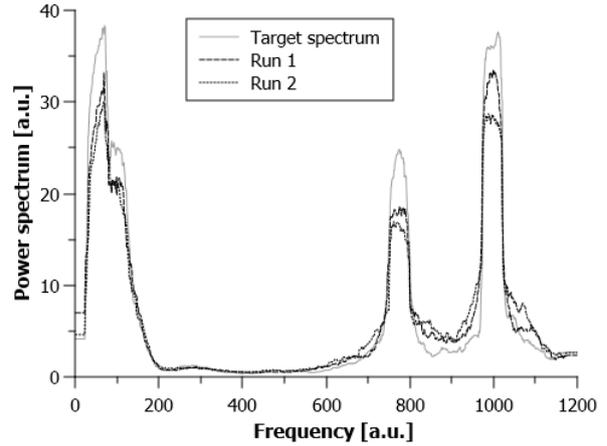

**Fig. 6.** Target spectrum and the starting spectra of two equivalent but different realizations of evolution (the only difference is the initial spatial and current state of the network). The target spectrum is the spectrum that the network output (3 of 10 neurons from the network) fits to during the evolution. The starting spectra are the spectra of network output when the network is in its initial spatial state. All parameters are set as described in the text (Sec. 4).

**Fig. 7.** Target spectrum and the spectra of the evolved networks (two separate runs of evolution). The resulting spectra of Run 1 and Run 2 are the output spectra of networks that fitted to the target spectrum, therefore the evolution finished with success. All parameters were set as described in the text (Sec. 4).

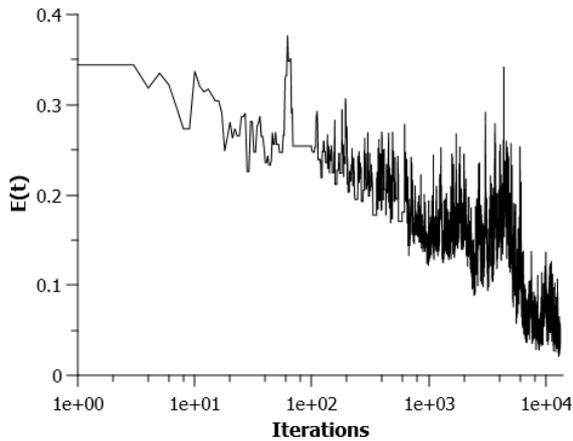 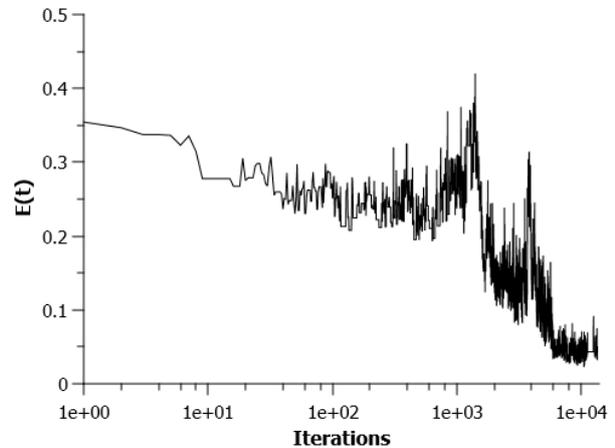

**Fig. 8.** Target functions $E(t)$ of the networks from Fig. 6 and 7 during the evolution. Horizontal axis is presented in logarithmic scale. The system reaches the spatio–current state of the network that fits the target function but the time needed for the system to evolve is up to 10000 iterations of Metropolis–algorithm evolution. All of the parameters were set as described in the text (Sec. 4).

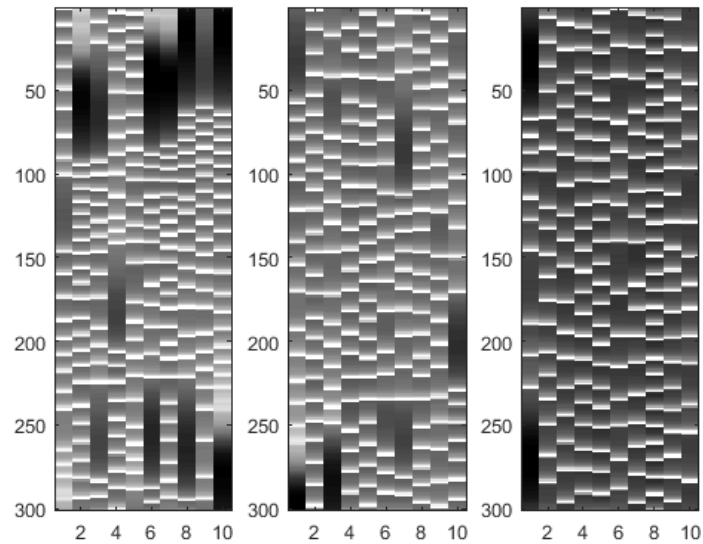

**Fig. 9.** Spatiotemporal patterns generated by all of the neurons in the network at three stages of the evolution: at the beginning (left), after 500 iterations (center) and after 10000 iterations (right). Horizontal axis presents the number of the node, vertical axis is time of the single simulation (the last part). During the whole process of evolution the system stays in the desynchronized state. The output nodes are #1, #4 and #7. At the beginning of the evolution all neurons present bursting chaotic behavior. When the system evolves, neurons #4 and #7 change their dynamics to regular periodic.

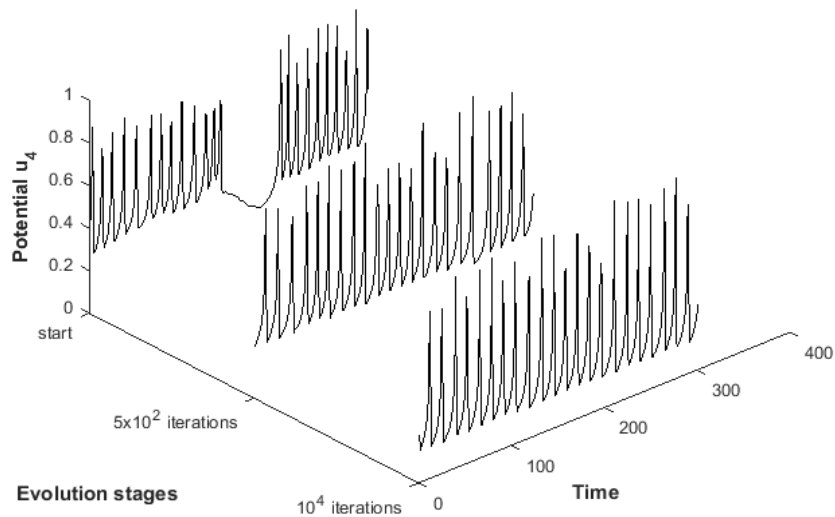

**Fig. 10.** Action potentials of one of the output neurons (neuron #4) changing during the evolution. At the beginning of the evolution the neuron presents bursting chaotic behavior. As the evolution goes on, the dynamics of the neuron changes to roughly periodic (500 evolution iterations) and to regular periodic (here, plotted after 10000 evolution iterations) with the period corresponding to one of the target spectrum peaks.

Assuming the same target spectrum but starting from the widely–spanned network (e.g. $x_{start}, y_{start} \in [-10, +10]$), we observed that the evolution converges to a positive result (Fig. 11) and resulting evolved network spatial property (mean length of the 3 shortest links) is similar to the one that started from the narrow–spanned group of neurons (as shown in Fig. 12). Taking into consideration Fig. 3, that result implies that the network with time delays of specific pre–assumed output properties has a specific spatial dimensions that allows it to function.

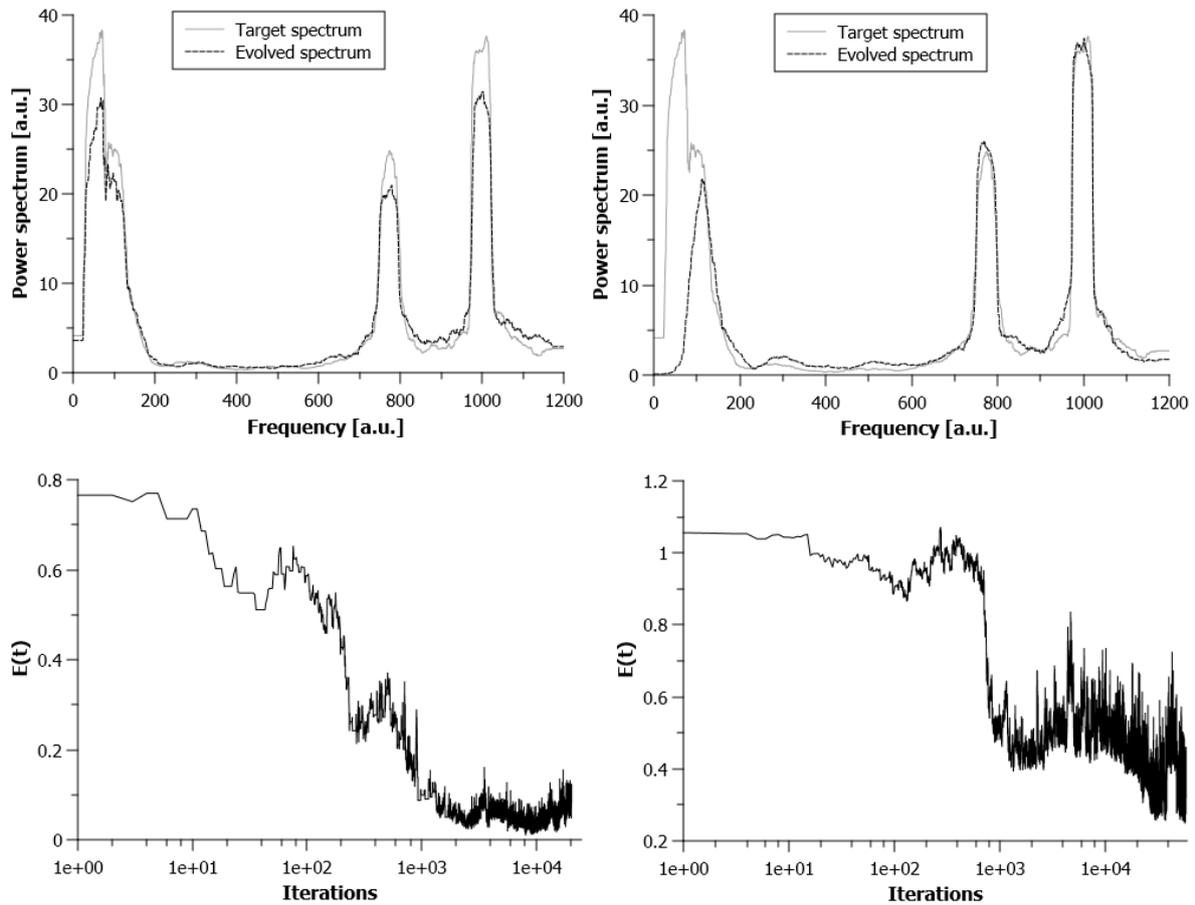

**Fig. 11.** Spectra and $E(t)$ diagrams of the networks evolved from the widely–spanned initial spatial conditions ($x_{start}, y_{start} \in [-10, +10]$ – left diagrams; $x_{start}, y_{start} \in [-50, +50]$ – right diagrams). It can be noted that the chaotic peak on the right–hand spectrum diagram is not completely fulfilled. If initial conditions are more sparse the evolution is less likely to converge. All of the parameters were set as described in the text (Sec. 4).

To check whether our study is robust to the network size we performed the simulation covering network of 20 neurons with 6 output nodes constructed as in the previous case. One of the resulting power spectra is presented in Fig. 13 and indicates that increasing the network size generates consistent results and that during the spatio–current evolution described here any target spectrum can be achieved (i.e. unlimited number of peaks, unrestricted magnitude of both chaotic peak and periodic peaks) if only sufficient number of output neurons and suitable size of the network provided.

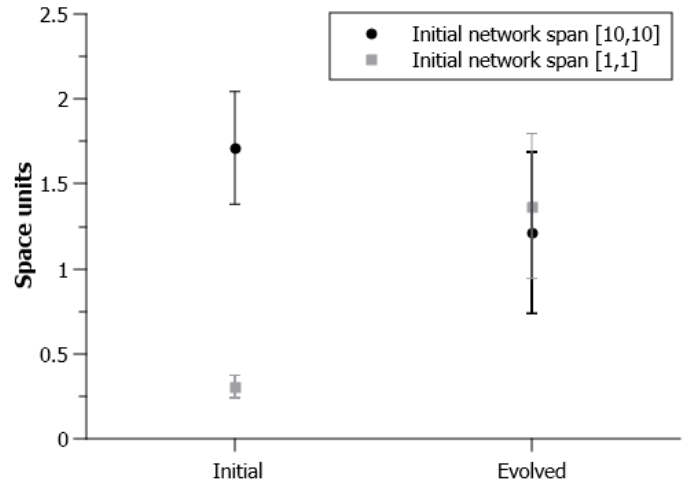

**Fig. 12.** Average length of the 3 shortest links in the network, averaged over 5 evolution runs. Both, initially widely–spanned and initially narrow–spanned networks tend to have similar value of this parameter if evolved positively (the resulting spectrum matches the target spectrum). Error bars represent standard deviation. This result indicates that there is a specific minimal physical size of the network to realize the target output spectrum.

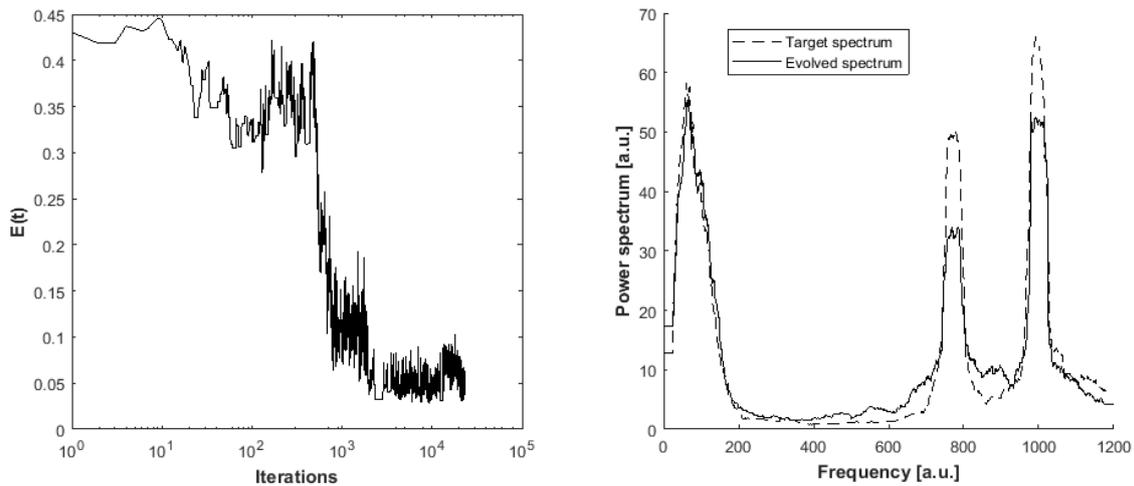

**Fig. 13.** $E(t)$ diagram and the spectrum of the network of 20 neurons evolved in the same way as the previous examples with 10 neurons. The target spectrum consists of one chaotic and two periodic peaks. Each peak is the result of the output signal from 2 output neurons. The success of the evolution of the network with 20 nodes indicates that the network and evolution models are robust to the network size.

## 5. Discussion and conclusions

In this paper, we have investigated the phenomenon of spatial evolution of the Hindmarsh–Rose neural network with time delays depending on the arrangement of neurons on the physical surface. It has been shown that under the specific assumptions and using appropriate parameters the neural evolution can converge to the desirable state, despite the simplicity of the target function and evolution model, and a large variety of spectra can be achieved. This is an interesting phenomenon as it is known that periodic and chaotic stimuli can change behavior of the neurons that are stimulated and the "control" network generating a large variety of signals can be evolved using a simple physical process. We have clearly shown that there is a specific range of network size in space which allows it to generate pre–assumed output. If this size is not reached the output dynamics is always chaotic. In turn, if all of the links exceeds the threshold length then the network loses its ability to produce chaotic signals. If links are very long periodic dynamics of output neurons also degenerates. It can be concluded that a network (or a system, generally speaking) with time delays (corresponding to physical space) of specific pre–assumed output properties has a specific spatial dimension that allows it to function. It is in compliance with earlier results where different phenomena like intermittent behavior or spiral waves were observed due to the specific set of conditions including proper time delays [19, 22] but here, in contrast, the network evolved itself fitting its output to the target. The successful use of Metropolis–Hastings Monte Carlo algorithm to evolve the system indicates that this approach can be used to search for specific states in the multidimensional parameter space of neural networks.

There are many possible extensions to our work. The networks we simulated were small due to the huge computational load. Varying parameters as driving current $I_i$ or coupling strength $g_{ij}$ can provide further interesting information. It is also possible to apply another target function or more complex dynamics to the model. Ultimately, modeling evolution of larger networks with small world connectivity matrices with shortcuts and clusterization would be an issue of the highest interest.

We hope our study will contribute to the understanding of complex behavior in population of interacting neurons.

## Acknowledgments

The work has been supported from the National Science Centre in Poland (grant no. 2012/05/E/ST2/02300).

**References**

1. Jalili, M.: Synchronizing Hindmarsh–Rose neurons over Newman–Watts networks. Chaos **19**, 033103 (2009)
2. Belykh, I., de Lange, E., Hasler, M.: Synchronization of Bursting Neurons: What Matters in the Network Topology. Phys. Rev. Lett. **94**, 188101 (2005)
3. del Molino, L.C.G., Pakdaman, K., Touboul, J., et al.: Synchronization in random balanced networks. Phys. Rev. E **88**, 042824 (2013)
4. Rosin, D.P., Rontani, D., Gauthier, D.J, et al.: Control of Synchronization Patterns in Neural-like Boolean Networks. Phys. Rev. Lett. **110**, 104102 (2013)
5. Leone M.J., Schurter, B.N, Letson, B., et al.: Synchronization properties of heterogeneous neuronal networks with mixed excitability type. Phys. Rev. E **91**, 032813 (2015)
6. Kwok, H.F., Jurica, P., Raffone, A.: Robust emergence of small–world structure in networks of spiking neurons. Cogn. Neurodyn. **1**, 39–51 (2007)
7. Dhamala, M., Jirsa, V.K., Ding, M.: Enhancement of Neural Synchrony by Time Delay. Phys. Rev. Lett. **92**, 074104 (2004)
8. Kinzel, W., Englert, A., Reents, G., et al.: Synchronization of networks of chaotic units with time–delayed couplings. Phys. Rev. E **79**, 056207 (2009)
9. Englert, A., Heiligenthal, S., Kinzel, W., et al.: Synchronization of chaotic networks with time–delayed couplings: An analytic study. Phys. Rev. E **83**, 046222 (2011)
10. Tang, G., Xu, K., Jiang, L.: Synchronization in a chaotic neural network with time delay depending on the spatial distance between neurons. Phys. Rev. E 84, 046207 (2011)
11. Zhu, J., Chen, Z., Liu, X.: Effects of distance–dependent delay on small–world neuronal networks. Phys. Rev. E **93**, 042417 (2016)
12. Zhang, J., Wang, C., Wang, M., et al.: Firing patterns transition induced by system size in coupled Hindmarsh–Rose neural system. Neurocomputing **74**, 2961–2966 (2011)
13. Wang, H., Ma, J., Chen, Y., et al.: Effect of an autapse on the firing pattern transition in a bursting neuron. Commun. Nonlinear Sci. Numer. Simulat. **19**, 3242–3254 (2014)
14. Wang, G., Jin, W., Wang, A.: Synchronous firing patterns and transitions in small-world neuronal network. Nonlinear Dyn. **81**, 1453–1458 (2015)
15. Ma, J., Qin, H., Song, X., Chu, R.: Pattern selection in neuronal network driven by electric autapses with diversity in time delays. Int J Mod Phys B **29** 1450239 (2015)
16. Ma, J., Tang, J.: A review for dynamics of collective behaviors of network of neurons. Sci China Technol Sci **58**, 2038–2045 (2016)
17. Ma, J., Tang, J.: A review for dynamics in neuron and neuronal network. Nonlinear Dyn. **89**: 1569-1578 (2017)
18. Wang, Q., Duan, Z., Perc, M., Chen, G.: Synchronization transitions on small-world neuronal networks: Effects of information transmission delay and rewiring probability. EPL **83**, 50008 (2008)
19. Wang, Q., Duan, Z., Perc, M., Chen, G.: Synchronization transitions on scale-free neuronal networks due to finite information transmission delays. Phys. Rev. E **80**, 026206 (2009)
20. Wang, Q., Duan, Z., Perc, M., Chen, G.: Impact of delays and rewiring on the dynamics of small-world neuronal networks with two types of coupling. Physica A **389**, 3299–3306 (2010)
21. Guo, D., Wang, Q., Perc, M.: Complex synchronous behavior in interneuronal networks with delayed inhibitory and fast electrical synapses. Phys. Rev. E **85**, 061905 (2012)
22. Wang, Q., Duan, Z., Perc, M., Chen, G.: Delay-enhanced coherence of spiral waves in noisy Hodgkin-Huxley neuronal networks. Phys. Lett. A **372**, 5681-5687 (2008)
23. Sun, X., Perc, M., Kurths, J.: Effects of partial time delays on phase synchronization in Watts-Strogatz small-world neuronal networks. Chaos **27**, 053113 (2017)
24. Monasson, R., Rosay, S.: Transitions between Spatial Attractors in Place-Cell Models. Phys. Rev. Lett. **115**, 098101 (2015)

25. Falahian, R., Dastjerdi, M.M., Molaie, M., et al.: Artificial neural network-based modeling of brain response to flicker light. Nonlinear Dyn. **81**, 1951–1967 (2015)
26. Purves, D., Augustine, G.J., Fitzpatrick, D., et al., editors: Neuroscience. 2nd edition. Sinauer Associates (2001)
27. Hastings, W.K.: Monte Carlo sampling methods using Markov chains and their applications. Biometrika **57**(1), 97–109 (1970)
28. Hindmarsh, J.L., Rose, R.M.: A Model of Neuronal Bursting Using Three Coupled First Order Differential Equations. Proc. R. Soc. Lond. Ser. B. Biol. Sci. **221**, 87–102 (1984)
29. Deng, L.; Yu, D.: Deep Learning: Methods and Applications. Found. and Trends in Signal Proc. **7**(3–4), 197–387 (2014)
30. Gerstner, W., Kistler, W.M., Naud, R., et al.: Neuronal Dynamics: From single neurons to networks and models of cognition. Cambridge Univ. Press (2014)